\documentclass[doublecol]{epl2} 
\usepackage{amsmath}
\title{Geometry-induced flocking and topological sound on a defect-free curved surface}
\shorttitle{Geometry-induced flocking and topological sound on a defect-free curved surface} 

\author{Devendra Saini\inst{1}
\thanks{E-mail: \email{d.k.saini27022003@gmail.com}} 
\and Atanu Bhatta\inst{2}\thanks{E-mail: \email{atanu.bhatta@dituniversity.edu.in}} \and Pritha Dolai\inst{1}\thanks{E-mail: \email{pritha@nitk.edu.in}}}
\shortauthor{D. Saini \etal}

\institute{                    
  \inst{1} Department of Physics, National Institute of Technology Karnataka, Surathkal, Mangalore-575025, India\\
  \inst{2} Department of Physics, School of Physical Sciences, DIT University, Dehradun- 248009, India
}

\abstract{
We study an ordered polar active flock on a torus and show that topological sound persists on a compact curved surface without topological defects or physical boundaries. Using the covariant Toner–Tu theory, we derive an effective non-Hermitian Dirac operator whose curvature-induced mass changes sign across the outer and inner equators, producing two Jackiw–Rebbi domain walls. These support co-propagating but distinct chiral edge excitations: a density mode localized on the positively curved outer equator and a Goldstone mode localized on the negatively curved inner equator. The bulk bands possess opposite half-integer Chern numbers whose jumps across the domain walls are determined by the sign of the Gaussian curvature. We further show that the localized modes are protected by a one-dimensional Callias index theorem, while the sum of the local indices obeys the Poincar\'e–Hopf theorem on the compact surface. Our results establish that curvature alone, independent of defects and boundaries, is sufficient to generate and protect topological sound in active matter, providing a unified connection between non-Hermitian topology, differential geometry, and hydrodynamic theory of collective motion.
}

\begin{document}

\maketitle

\section{Introduction}
Flocking, the self-organized collective motion of active, self-propelled units, is a paradigmatic example of non-equilibrium system~\cite{vicsek1995,marchetti2013}. A defining feature of non-equilibrium ordered state of polar active matter is the breaking of time-reversal symmetry~\cite{bowick2022}. When such a flock placed on a curved substrate, its orientational order is frustrated by the geometry, and supports long-wavelength sound modes
that are gapped by the curvature, producing a band structure with nontrivial topology~\cite{fily2016, shankar2017, tubiana2024}. The result is analogous to the edge states of quantum Hall systems and of engineered topological metamaterials~\cite{sone2019,Kane2014,Souslov2017,Nash2015,Delplace2017,zhang2018,sone2026,palma2021,shankar2022}. A topological fluctuation theorem governed by vortex winding numbers has also been established for stochastic systems~\cite{tang2021,mahault2022}. Collective motion on curved surfaces is moreover ubiquitous in biology, from morphogenesis to tissue repair~\cite{streichan2017,ewald2008,Yevick2015,castrovillarreal2018,schamberger2023,brandstatter2023}.

However, it is demonstrated explicitly in~\cite{shankar2017} that on a sphere a density edge mode localizes on the equator, whereas on the catenoid a Goldstone mode localizes on the neck. On the sphere, the topological sound is entangled with defects and one cannot rule out that these defects, rather than the local curvature, are responsible for the protected modes~\cite{sknepnek2015}. The catenoid removes the defects but is
noncompact, so its single edge mode coexists with physical boundaries whose role is hard to disentangle. It is therefore natural to ask: \emph{are the topologically protected modes a consequence of the local geometry of the surface, independent of defects and boundaries?}

Here we answer this affirmatively by studying an active polar flock on a torus. The dynamics of single active Brownian particles on a torus has been formulated in a Riemannian framework~\cite{apaza2018}. Experimentally, toroidal droplets and toroidal nematic shells have been realized and are stable over controlled lifetimes~\cite{Pairam2013}, and active suspensions confined to curved interfaces~\cite{Keber2014,pearce2019,pearce2020} have also been studied. Related active droplets can exhibit defect mediated spontaneous rotation~\cite{nejad2023}.

Torus is a closed surface with Euler characteristic $\chi=0$, so the Poincar\'e--Hopf theorem imposes \emph{no} defects. The ordered flock is a globally smooth azimuthal band with vanishing defect charge~\cite{nambisan2024}. Nevertheless, we find that topological sound persists, establishing that it is a property of the local geometry. The Gaussian curvature of the torus changes sign, positive on the outer half, negative on the inner. So the gap-closing condition is met on \emph{two} geodesics, the outer and inner equators. Thus they behave like domain walls, binding protected modes of \emph{opposite} character, a density wave on the positively curved outer equator and a Goldstone wave on the negatively curved inner equator, co-rotating at independently tunable speeds. The torus is thus the minimal closed, defect-free surface carrying both type of edge-modes simultaneously.

Remarkably, the Dirac operator at each gap-closing geodesic is a Fredholm operator whose index is computed by the one-dimensional Callias index theorem~\cite{callias1978}, and we show that this index equals the sign of the Gaussian curvature at the geodesic. Summing over all walls of a closed surface of revolution returns the Euler characteristic. This ties the momentum-space edge-mode count to the real-space topology through a Gauss-Bonnet type relation. We also find that the aspect ratio (ratio of the toroidal to the poloidal radius) of the torus acts as a single control parameter interpolating between a sharply localized two-channel waveguide (fat-ring limit) and the gapless cylinder ($R/r\to\infty$, where the Gaussian curvature vanishes).

The paper is organized as follows. In the first section we set up the covariant Toner-Tu model on the torus. Next section obtains the defect-free ordered steady state. Then, the Toner-Tu equations are linearized about the steady state and the curvature-gapped sound spectrum is obtained in the third section. After that we compute the band Chern numbers and find its relation to the curvature topology. Next, we calculate the edge-states explicitly . Finally, we formulate the index principles responsible for the topologically protected modes localized at the outer and inner equator, discuss the results and conclude.
\section{Toner-Tu Equations on Torus}
\label{sec:model}
Toner-Tu hydrodynamics of an overdamped polar active fluid with density $\rho$ and the polarization density $\mathbf{p} $ is described by the continuity equation~\cite{tonertu1995,tonertu1998,toner2005,geyer2018}
\begin{equation} \label{eq:continuity}
    \partial_t \rho + \nabla_\mu p^\mu = 0
\end{equation}
where $\nabla_\mu p^\nu  = \partial_\mu p^\nu + \Gamma^\nu_{\alpha\mu} p^\alpha$ with Christoffel symbols  $\Gamma^\nu_{\alpha\mu}$ and the polarization obeys
\begin{align} \label{eq:toner_tu}
    \partial_t p^\mu + \lambda p^\mu \nabla_\nu p^\nu  &= [a (\rho - \rho_c) - b g_{\alpha\beta}p^\alpha p^\beta]p^\mu \nonumber \\
    &\!\! + \nu (\Delta p^\mu + K_G p^\mu) + \nu' \nabla^\mu \nabla_\nu p^\nu - v_1 \nabla^\mu \rho \nonumber \\
\end{align}
where $\nu$ and $\nu'$ are the shear and bulk viscosities respectively, $\lambda$ is a kinematic convective parameter, $v_1 (>0)$ is a compressional modulus, and $a, b (>0)$ are the parameters that set magnitude of the mean-field polarized state for $\rho > \rho_c$ with the critical density $\rho_c$ for the flocking transition. 

We parameterize the torus by the poloidal angle $\theta \in [0, 2\pi)$ and toroidal angle $\phi \in [0, 2\pi) $, with the tube radius $r$ and center-to-tube distance $R>r$. The metric on the torus $(T^2)$ is 
\begin{equation}
    ds^2= g_{\alpha\beta} dx^\alpha dx^\beta  = r^2 d\theta^2 + (R + r  \cos \theta)^2 d\phi^2  
\end{equation}
The Gaussian curvature is 
\begin{equation}
    K_G  = \frac{\cos \theta}{r^2 (\varepsilon + \cos \theta)}; \qquad \varepsilon  = \frac{R}{r}
\end{equation}
The Gaussian curvature is positive on the outer half of the tube $|\theta| \le \pi/2$ and negative on the inner half $(\pi/2 < \theta < 3\pi/2)$, vanishing on the top and bottom circles $\theta = \pm \pi/2$.

The non-vanishing Christoffel symbols are
\begin{equation}
    \Gamma^\theta_{\phi\phi} = \delta (\theta) \sin \theta, \quad \Gamma^\phi_{\theta\phi} = - \sin \theta/\delta (\theta); \quad \delta = (\varepsilon + \cos \theta) \nonumber 
\end{equation}
%
%
\section{Defect-Free Ordered Steady State}
Since the Euler characteristic $\chi (T^2) = 0$, the oriented order on the torus requires no defects, a smooth non-vanishing vector field exists (e.g. $\partial_\phi $). Accordingly, we seek an order steady state $\rho = \rho_{ss} (\theta), p^\theta =0, p^\phi = p^\phi_{ss} (\theta)$ having azimuthal symmetry, a band of flow circulating in the toroidal direction. Viscous terms are neglected as their contribution is subdominant compared to the contribution coming from activity.

The steady state satisfies the continuity equation~\eqref{eq:continuity} identically. Neglecting the higher derivative terms, the $\theta$ and $\phi$ components of~\eqref{eq:toner_tu} reads
\begin{align} \label{eq:theta}
    \lambda \, \delta \, \sin \theta \, (p^\phi_{ss})^2 &= - (v_1/r^2) \, \partial_\theta \rho_{ss} \\ \label{eq:phi}
    p^\phi_{ss}[a(\rho_{ss}- \rho_c)-b r^2 \delta^2 (p^\phi_{ss})^2] &= 0 
\end{align}
The $\phi$-equation~\eqref{eq:phi} fixes the magnitude, $b\, r^2\, \delta^2 \, (p^\phi_{ss})^2 = a (\rho_{ss} - \rho_c)$. Substituting this in~\eqref{eq:theta} we get
\begin{equation}
    \rho_{ss} (\theta) = \rho_c + (\rho_{max} - \rho_c) \left[ \delta(\theta)/\delta_{max} \right]^\eta 
\end{equation}
where $\eta = \lambda a/(b v_1) (> 0)$, $ \delta_{max} = \delta(0)$, and $\rho_{max} = \rho_{ss} (0)$. Fixing $\rho_{max}$ in terms of the mean density $\rho_0 = \langle \rho_{ss} \rangle $ over the torus, with area element $dA = r^2 \, \delta (\theta)\,d\theta d\phi $ and total area $A = 4\pi^2 Rr$, we get
\begin{equation} 
\label{eq:rho_ss}
    \rho_{ss} (\theta) = \rho_c + (\rho_0 - \rho_c) \, B_\eta \, [\delta(\theta)]^\eta
\end{equation}
where
\begin{equation}
    B_\eta  = \frac{ 2\pi \varepsilon}{\int_0^{2\pi} [\delta (\theta)]^{\eta + 1} d\theta}.
\end{equation}
\begin{figure}
    \centering
\includegraphics[width=0.7\linewidth]{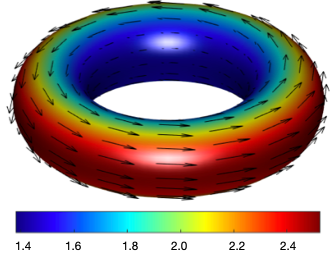}    
\caption{Steady state density profile and polarization on a torus for $\eta$=2\,. The color shows maximum density (red) at the outer equator and minimum density (blue) at the inner equator. The arrow indicates the direction of polarization.
}
\label{fig:density_torus}
\end{figure}

\begin{figure}
\includegraphics[width=\linewidth]{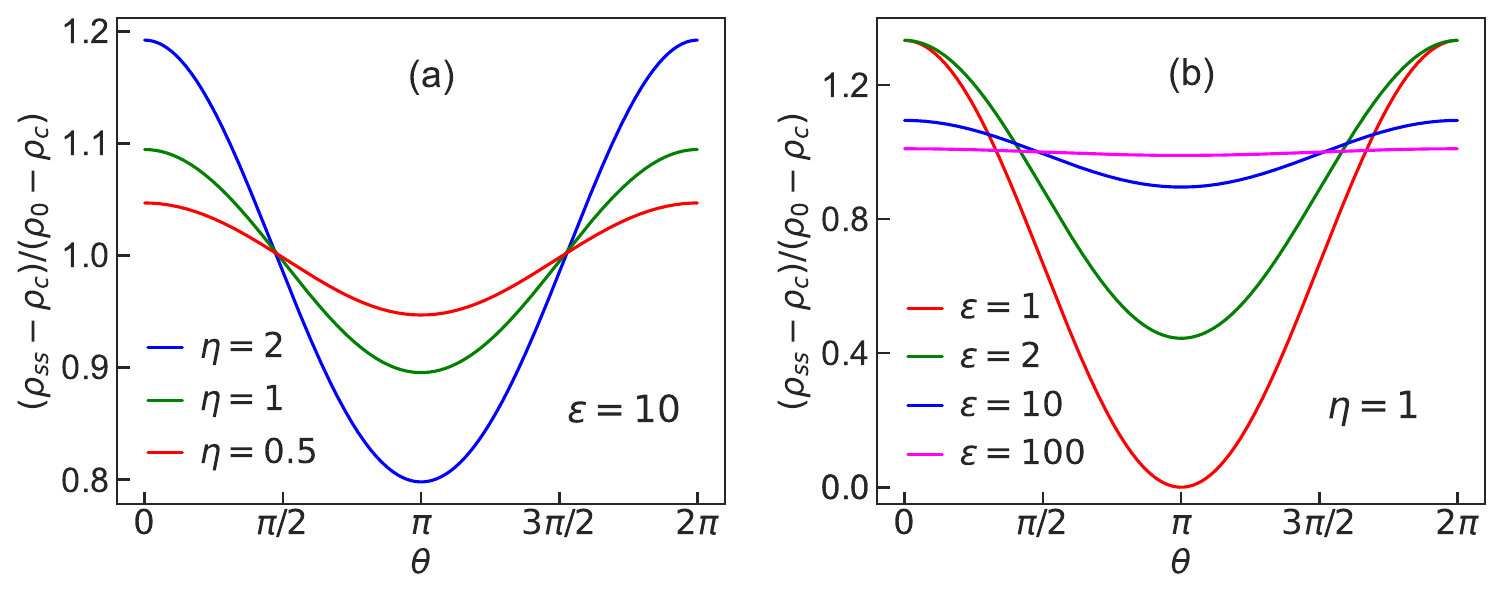}
\caption{Normalized steady state density profile as a function of poloidal angle $\theta$ (a) Variation with $\eta$ at fixed aspect ratio $\varepsilon=10$. (b) Variation with aspect ratio $\varepsilon$ at a fixed value of $\eta=1$.}
\label{fig:rho_ss-vs.-theta}
\end{figure}

Fig.~\ref{fig:density_torus} shows the steady state density profile on a torus. The arrows indicate the polarization direction which is chosen spontaneously.  
The ordered state exists for $\rho_0 > \rho_c$.  and the density is maximum ($\rho_{\rm max}$) on the outer equator $\theta = 0 $ and minimum ($\rho_{\rm min}$) on the inner equator $\theta = \pi $ (Fig.~\ref{fig:rho_ss-vs.-theta}). Self-advection lenses the flow along geodesics and crowds the flock onto the outer rim. There are no vortices, the profile is smooth and nodeless everywhere on the compact tube~\cite{chardac2021}.

\begin{figure}[h]
\centering
\includegraphics[width=0.75\linewidth]{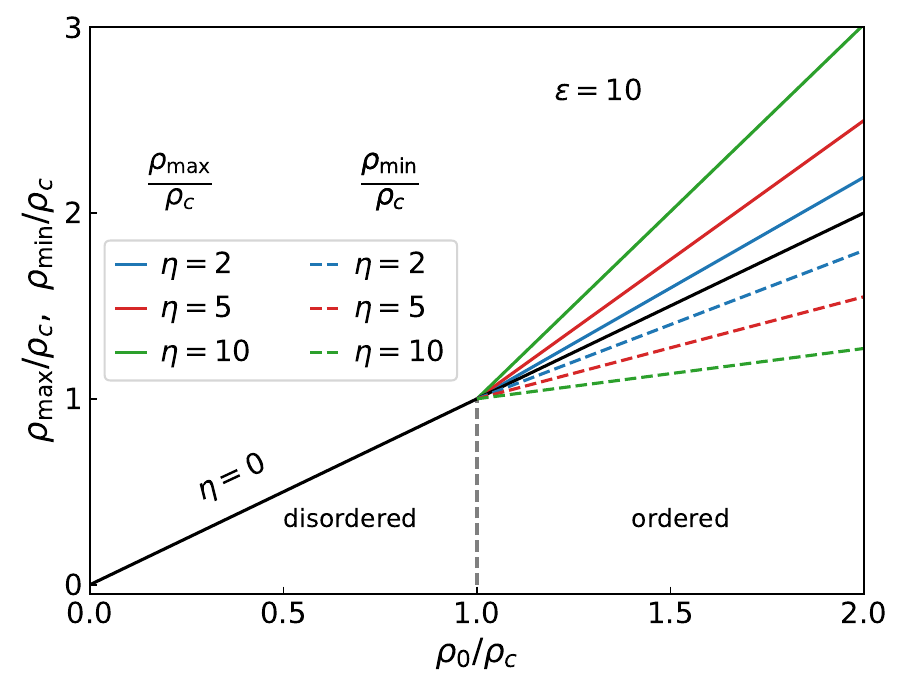}
\caption{Maximum density ($\rho_{\rm max}$) and minimum density ($\rho_{\rm min}$) plotted  as a function of mean density $\rho_0$ for different values of $\eta$. The solid lines indicate maximum density and the dotted lines show minimum density. The system remains in the disordered phase for $\rho_0 < \rho_c$ as expected. There exist a polar band for $\rho_0 > \rho_c$. The black solid line corresponds to $\eta=0$ and resulting in $B_{\eta}=1$, indicating homogeneous density profile in flat space. }
\label{fig:ordered_braches}
\end{figure}

The outer-equator peak grows linearly with $\rho_0$ as
\begin{equation}
    \rho_{\rm max} = \rho_c + s_{\rm out} \, (\rho_0 -\rho_c), \quad s_{\rm out} = B_\eta (\varepsilon + 1)^\eta 
\end{equation}
while the inner-equator trough grows more slowly 
\begin{equation}
    \rho_{\rm min} = \rho_c + s_{\rm in} \, (\rho_0 -\rho_c), \quad s_{\rm in} = B_\eta (\varepsilon - 1)^\eta 
\end{equation}
with $s_{\rm in}/s_{\rm out} = [(\varepsilon -1)/(\varepsilon + 1)]^\eta < 1$. The two branches bracket the mean-density line, which both meet at $(\rho_c, \rho_c) $ (Fig.~\ref{fig:ordered_braches}). Extrapolated to $\rho_0 = 0$, the ordered branches have a negative intercept set by the aspect ratio $\varepsilon$. It's worthwhile here to mention that, whereas for the sphere the slope depends only on $\eta$, and is independent of the size of the sphere, for the torus the slopes depend on the aspect ratio $R/r$.

Nevertheless, the torus slopes approach unity in the flat-cylinder limit $(R/r \to \infty)$, a nearly homogeneous flock. They spread apart as $R \to r$ (Fig.~\ref{fig:ordered_braches}). This aspect ratio dependence is a direct and measurable signature of the toroidal geometry.
\section{Linearized Dynamics and Gapped Spectrum}
Since the curvature does not affect the disordered phase~\cite{shankar2017}, we linearize about the ordered state $\rho = \rho_{ss} + \delta \rho$, $p^\mu = p_{ss}^\mu + \delta p^\mu$, in a tangent plane at a reference latitude $\theta_0$. We consider the tangent space coordinates $(x,y)$ as $\theta = \theta_0 + y/r, \phi = x /r\delta_0$, where $ \delta(\theta_0) = \delta_0$. Now, writing $\delta p^\theta = v,\, \delta p^\phi = u$, we perform a Galilean boost $x \to x -\lambda p_0 t $ to the frame comoving with the longitudinal mode, where $p_0 = p^\phi_{ss}(\theta_0)$. Then the linearized continuity equation and the Tone-Tu equations read
\begin{align}
    \partial_t \delta\rho &= \bar \lambda \partial_x \delta \rho - \partial_x u - \partial_y v + m_0 v  \\
    \partial_t u &= - \frac{\bar v_1}{(\varepsilon + \cos \theta_0)^2} \partial_x \delta \rho + \alpha \, \delta \rho - \beta \, u + \frac{\bar \lambda m_0 }{2} (\eta + 2) v \\
    \partial_t v &= - \bar v_1 \partial_y \delta \rho - \frac{2 \bar \lambda}{m_0} \sin \theta_0 u 
\end{align}
with the redefined parameters given in the Table~\ref{tab:redefined_param} and 
\begin{equation}
    m(\theta) = \sin \theta/\delta(\theta); \qquad m(\theta_0) = m_0
\end{equation}
\begin{table}[h!]
\begin{center}
\begin{tabular}{cccc}
\hline
\hline
$\lambda$  & $a$ & $b$ & $v_1$ \\
\hline 
$\bar \lambda =\lambda p_0 $ & $\alpha =  a p_0$ & $\beta = 2 b p_0^2 r^2 \delta^2$  &  $\bar v_1 = \frac{v_1}{r^2}$ \\
\hline 
\hline 
\end{tabular}
\end{center}
\caption{Toroidal parameters} 
\label{tab:redefined_param}
\end{table}

Finally we recast the above coupled linear equations as follows 
\begin{equation}
    \partial_t |\Psi \rangle = H |\Psi\rangle; \qquad |\Psi\rangle = \begin{bmatrix} \delta\rho & v  & u\end{bmatrix}^T
\end{equation}
In the Fourier space defined as $\tilde{\Psi} (\mathbf{q)} = \int d^2\mathbf{r} ~e^{-i \mathbf{q \cdot r}} \Psi(\mathbf{r})$, $H$ becomes
\begin{equation}
    H (\mathbf{q}) = \begin{bmatrix}
                        - \bar \lambda q_x & q_x & q_y + i m_0 \\
                        i \alpha + \frac{v_1}{\delta_0^2} q_x & - i \beta & i \frac{\bar \lambda m_0}{2} (\eta +2) \\
                        \bar v_1 q_y &  - 2i \frac{\bar \lambda}{m_0^2} \sin^2\theta_0 & 0
                     \end{bmatrix}
\end{equation}

\begin{figure*}[th]
    \centering \includegraphics[width=0.9\linewidth]{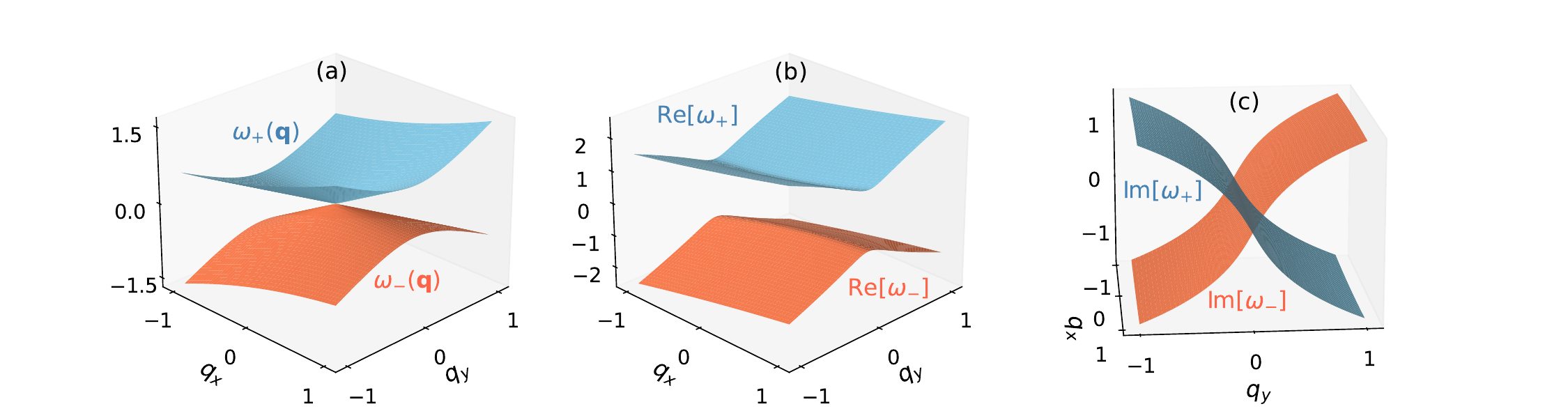}
\caption{Eigenfrequencies plotted as a function of wave vector. (a) $m=0$, (b) Real part of $\omega_{\pm}({\bf q})$ for $m=0.1$, (c) Imaginary part of  $\omega_{\pm}({\bf q})$. Here, $R$=3, $r$=1, $\theta_0=23.1^o$, $m=0.1$, $\delta=3.92$, $\alpha=2$, $\beta=1$, $\bar{\lambda}=1$, $\bar{v_1}=1$.}
\label{fig:omega_plot}
\end{figure*}

Note that, on the outer and inner equators ($\theta_0 = 0$ and $\pi$ respectively), $m_0 = 0$. In the vicinity of the outer and inner equator where $m_0$ is non-zero but very small, the dispersion relations read 
\begin{align}
    i \omega_0 &=  \beta - i \frac{\alpha}{\beta} \left( q_x  + 2 m\bar \lambda \delta_0 q_y \right) + \mathcal{O} (q^2, m^2) \\
    \omega_\pm &= \frac{1}{2\beta} (\alpha - \bar \lambda \beta) q_x  \nonumber \\ 
    & \!\!\!\!\!\!\pm \left[ (\alpha - \bar \lambda \beta)^2 q_x^2 + \!4 \beta \left(\frac{m_0}{\delta_0}\! - \!i q_y\right)(2 \alpha \bar \lambda m_0 \delta_0 + i \beta \bar v_1 q_y \right]^{\frac{1}{2}}\nonumber \\ & \qquad \qquad \qquad \qquad \qquad \qquad \qquad+ \mathcal{O} (q^2, m^2)
\end{align}
Near the outer equator ($\theta_0 = 0$), $m_0 =  0$ and the collective modes propagate symmetrically along the toroidal surface. The dispersion relations exhibits two propagating branches corresponding to the gapless hydrodynamic modes $\omega_\pm$ (Fig.~\ref{fig:omega_plot}(a)). The real part of the spectrum increases approximately linearly with the wave vector, indicating propagating sound-like modes in the ordered phase. The dynamics resemble those of an active fluid on flat space - as expected since the curvature vanishes at the outer equator. For $m_0 \neq 0$, the propagation becomes anisotropic due to curvature-induced coupling between the hydrodynamic modes as shown in Fig.~\ref{fig:omega_plot}(b) and \ref{fig:omega_plot}(c).    
\section{Band Topology and Curvature Topology}
The fluctuation $u$ of the polarization magnitude is a fast mode decaying on the microscopic time scale $\beta^{-1}$. On time $t \gg \beta^{-1}$ it is slaved to the density, $u \approx \alpha \, \delta/\beta$, and the slow dynamics reduces to $(\delta\rho , \, v)$
\begin{equation}
     \label{eq:dirac}
    \partial_t \begin{bmatrix} \delta\rho \\ v \end{bmatrix}  = D( \mathbf{q}) \begin{bmatrix} \delta\rho \\ v \end{bmatrix}
\end{equation}
where after rescaling $|\alpha-\beta \bar{\lambda}|q_x/\beta$ to $q_x$ to absorb the longitudinal speed $D(\mathbf{q})$ becomes
\begin{equation}
    \label{eq:Dq}
    D(\mathbf{q}) = \begin{bmatrix}
                        sq_x & q_y + i m \\
                        \bar v_1 q_y - i \mu m & 0
                    \end{bmatrix}
\end{equation}
with $s = \text{sgn}(\alpha - \bar \lambda \beta)$ and 
\begin{equation}
    \mu = \frac{2\alpha \bar \lambda \delta^2}{\beta} = \frac{a \lambda}{br^2} 
\end{equation}
Note that, the ratio $\mu/\bar v_1$ becomes the steady state exponent $\eta$, independent of the position.

The eigenvalues of $D(\mathbf{q})$ provide two slow sound branches. Note that, we can rewrite the Dirac matrix $D(\mathbf{q})$ as
\begin{equation}
    D(\mathbf{q}) = d_0 \mathrm{I}_{2\times2} + \boldsymbol{d} \cdot \boldsymbol{\sigma} 
\end{equation}
where $\mathrm{I}_{2\times2}$ is the two-dimensional identity matrix and $\sigma_i$'s are the Pauli matrices with $d_0 = sq_x/2$ and
\begin{equation}
    \boldsymbol{d} \cdot \boldsymbol{d} = \frac{s^2 q_x^2}{4} + \bar v_1 q_y^2 + \mu m^2 + i m q_y(\bar v_1 - \mu )
\end{equation}
The eigen-frequencies are given by
\begin{equation}
    \omega_\pm (\mathbf{q}) = \frac{1}{2} s q_x \pm \sqrt{\boldsymbol{d} \cdot \boldsymbol{d}}
\end{equation}
At $\mathbf{q} = 0$, the spectrum is gapped (Fig.~\ref{fig:omega_plot}(b))  by an amount proportional to $m$
\begin{equation}
 \label{eq:gap}
    \Delta (\theta)= |\omega_+(0) - \omega_-(0)| = 2 \sqrt \mu |m| = 2 \sqrt{\frac{a \lambda}{b}} \frac{\sin \theta}{R + r \cos \theta}\,.
\end{equation}
One can interpret the gap equation geometrically as it is a material speed $\sqrt{a\lambda/b}$ times the logarithmic gradient of the local circumference $ |\partial_{\ell} \ln \sqrt{g_{\phi\phi}}| $, where $\ell$ is arc length.

The gap closes (Fig.~\ref{fig:omega_plot}(a)) on outer $(\theta = 0)$ and inner $(\theta = \pi)$ equators, where $m = 0$.  In the the thin-cylinder limit $(\varepsilon \to \infty)$, $\Delta$ scales as $1/R$, yielding gapless spectrum for the cylinder.

However, for $m \neq 0$, the spectrum is gapped and each band carries a well-defined Berry structure. The bi-orthonormal eigenbasis of the non-Hermitian Dirac matrix $D(\mathbf{q})$ is constructed as
\begin{equation}
    D(\mathbf{q}) |\psi_\pm \rangle = \omega_\pm |\psi_\pm \rangle; \qquad \langle\chi_\pm| D(\mathbf{q}) = \langle \chi_\pm |\omega_\pm 
\end{equation}
where
\begin{equation}
    |\psi_\pm  \rangle = \begin{bmatrix}
        \frac{\omega_\pm}{\bar v_1 q_y- i \mu m } \,c_\pm \\ c_\pm
    \end{bmatrix}; \quad \langle \chi_\pm | = \begin{bmatrix}
        \frac{\omega_\pm}{q_y + i m} c_\pm & c_\pm 
    \end{bmatrix}
\end{equation}
with
\begin{equation}
    c_\pm = \left[1+ \frac{\omega_\pm^2}{(q_y +i m)(\bar v_1 q_y - i \mu m)} \right]^{-1/2}
\end{equation}
such that $\langle \chi_i  | \psi_j \rangle = \delta_{ij} ~( i,j = \pm)$.

The band topology can be studied by computing the abelian Berry connection and curvature~\cite{berry1984}
\begin{equation}
    \boldsymbol{\mathcal{A}}^\pm = i \langle \chi_\pm | \nabla_{\mathbf{q}} |\psi_\pm \rangle; \qquad  \mathcal{F}_\pm (\mathbf{q}) = \frac{\partial \mathcal{A}^\pm_y}{\partial q_x} -  \frac{\partial \mathcal{A}^\pm_x}{\partial q_y}\,.
\end{equation}
The explicit calculation yields 
\begin{equation}
\label{eq:berry_curvature}
    \mathcal{F}_\pm (\mathbf{q}) = \pm \frac{s\,m (\mu + \bar v_1)}{[s^2 q_x^2 + 4\mu m^2 + 4 \bar v_1 q_y^2 + 4 imq_y(\bar v_1-\mu)]^{3/2}}\,.
\end{equation}
Clearly, $\mathcal{F}_+ + \mathcal{F}_- = 0$ holds identically. Since, the imaginary part of~\eqref{eq:berry_curvature} is an odd function of $q_y$ (on principal branch), it integrates to zero. Finally, integrating over the non-compact $\mathbf{q}$-plane results half integer Chern number~\cite{nakahara2003}
\begin{equation}
    C_\pm = \int \frac{d^2 \mathbf{q}}{2\pi} ~\mathcal{F}_\pm (\mathbf{q)} = \pm \frac{s}{2} \text{sgn}(m)\,.
\end{equation}
The half-integer reflects the single Dirac cone structure of the continuum long-wavelength theory, the parity-anomaly contribution familiar in the studies of Chern insulators~\cite{haldane2015}. An ultraviolet regularization restores the integer quantization without affecting the difference of Chern numbers across a wall~\cite{hasan2010,ryu2010}. Therefore, we can use the bulk-edge correspondence to predict the topologically protected edge states on the outer and inner equator, where the gap closes. 

The mass $m(\theta)$ is related to the Gaussian curvature $K_G$ through as
\begin{equation} \label{eq:riccati}
    K_G (\theta) = \frac{m'(\theta) - m^2(\theta)}{r^2}; \qquad m'(\theta) = \frac{dm}{d\theta}\,.
\end{equation}
On the outer and inner equator, where $m = 0$, the above relation reduces to 
\begin{equation}
    K_G( \theta_*) = \frac{m'(\theta_*)}{r^2};  \qquad \theta_* = 0, \pi \,.
\end{equation}
So, the sense in which the mass crosses zero is fixed by the sign of the Gaussian curvature at the outer and inner equator. Consequently the change in the Chern number across the gap-closing geodesics becomes 
\begin{equation}
   \left. \Delta C_+ \right|_{\theta_*} = s \text{ sgn}(m'(\theta_*)) = s \text{ sgn} (K_G(\theta_*))\,.
\end{equation}
On the torus, $C_+(\theta_0)$ takes the value $+ s/2$ on the upper-half tube, and $-s/2$ on the lower-half tube. Crossing the outer equator $(K_G > 0)$ gives $\Delta C_+ = + s $, while crossing the inner equator $(K_G < 0)$ gives $\Delta C_+ = - s$. The net change around the tube vanishes 
\begin{equation}
    \sum \Delta C_+ = 0 = \frac{s}{2} \chi(T^2) 
\end{equation}
as it must hold for a closed genus $g$-surface with Euler characteristic $\chi = 2 - 2g$ and here the torus is a genus $g=1$ surface.  
\section{Topological Edge Modes}
As elucidated in~\cite{shankar2017}, each gap closing leads to a unidirectional localized edge mode. At each of the outer and inner equator when crossed the sign of the mass $m(\theta)$ changes, thus these geodesics behave as Jackiw-Rebbi domain walls that bind a chiral mode~\cite{jackiw1976, shankar2022}. Treating $q_x$ as a good quantum number and substituting $q_y \to -i \partial_y$, we seek a solution of~\eqref{eq:dirac} bound to the domain walls. Note that the mass term in the off-diagonal entries of $D(\mathbf{q})$ carries the opposite sign. So, near the domain walls, the components of the two-dimensional spinor solution acquire reciprocal exponential envelopes. Normalization condition on the compact tube selects only one of them yielding the chiral solution~\cite{leykam2017, sone2020}.

Now the equation~\eqref{eq:dirac} turns to two coupled first order differential equations
\begin{eqnarray}
\label{eq:diffop_A}
 -i \,\mathcal{L}_A \, \delta\rho &=&  \omega_e  v; \qquad \qquad \quad \mathcal{L}_A = \bar v_1 \partial_y + \mu m \\ \label{eq:diffop_B}
 i\, \mathcal{L}_B \, v &=& (sq_x - \omega_e) \,  \delta \rho; \quad \mathcal{L}_B = \partial_y - m
\end{eqnarray}
where $\omega_e$ is the eigen value of the differential operator $D (q_x, - i\partial_y)$. The first order differential operators $\mathcal{L}_A$ and $\mathcal{L}_B$ completely characterize the edge states at the domain walls. A topological edge state is the square-integrable kernel of the appropriate first order differential operator. The bases of the kernels of $\mathcal{L}_A$ and $\mathcal{L}_B$ are computed as 
\begin{equation}
    \delta\rho \sim  \exp\left[ - \frac{\mu}{\bar v_1}\int^y m(y') \,dy'\right]
\end{equation}
and
\begin{equation}
    v \sim  \exp\left[ \int^y m(y') \,dy'\right]
\end{equation}
respectively.
\subsection{Outer equator $(\theta = 0)$}
Here 
\begin{equation}
    m(y) \sim \frac{y}{R + r}\,.
\end{equation}
So, $m(y<0) <0$ and $m(y>0) > 0$, i.e., $m$ crosses the outer equator as $-\text{ve } \to +\text{ve}$. Hence
\begin{equation}
    \int^y m (y') \, dy' \sim + \infty \qquad \text{as} \qquad |y| \to \infty \,.
\end{equation}
Therefore, $\delta \rho$ is square-integrable while $v$ diverges as one goes far away from the outer equator on either side. Since, we are only interested in the square integrable bases of the kernels, $\text{ker}_{\mathcal{L}_A} \neq 0; \,\,\text{ker}_{\mathcal{L}_B} = 0\,.$
%
%
Then $\omega_e v = 0$. Now for a pure kernel state, $\omega_e \neq 0$; then from~\eqref{eq:diffop_A} $v=0$. Substituting in~\eqref{eq:diffop_B} we get $\omega_e = sq_x$. Finally, computing the integral
\begin{equation}
    \int_0^y m  \, dy' = r \int_0^\theta \frac{\sin \theta \,d\theta}{\varepsilon + \cos \theta}  = - r \ln \left( \frac{\varepsilon + \cos \theta}{\varepsilon + 1}\right)  \,, 
\end{equation}
we obtain the topological edge state at the outer equator as
%

\begin{equation}
    |\delta \Psi_{\rm edge} \rangle = \begin{bmatrix}
        \delta \rho_{\rm edge} \\ 0
    \end{bmatrix}
\end{equation}
with
\begin{equation}
    \delta \rho_{edge} = \left( \frac{\varepsilon + \cos \theta}{\varepsilon + 1}\right)^\eta \sum_{n\ge 0} [a_n e^{in (\phi - \frac{\alpha}{\beta}t)} + \text{c.c.}]\,.
\end{equation}
This is a chiral density mode circulating at angular velocity $\alpha/\beta$ with an envelope $\sim \delta^\eta $ which coincides with steady state band profile~\eqref{eq:rho_ss}.

\begin{figure}[t]
\centering

\includegraphics[width=0.48\columnwidth]{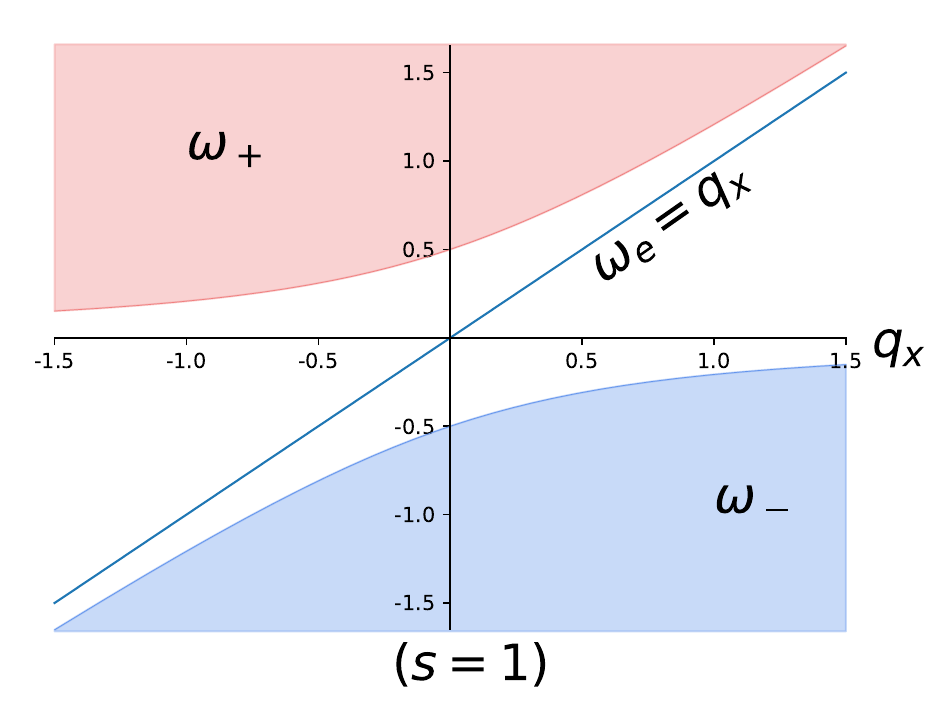}
\hfill
\includegraphics[width=0.48\columnwidth]{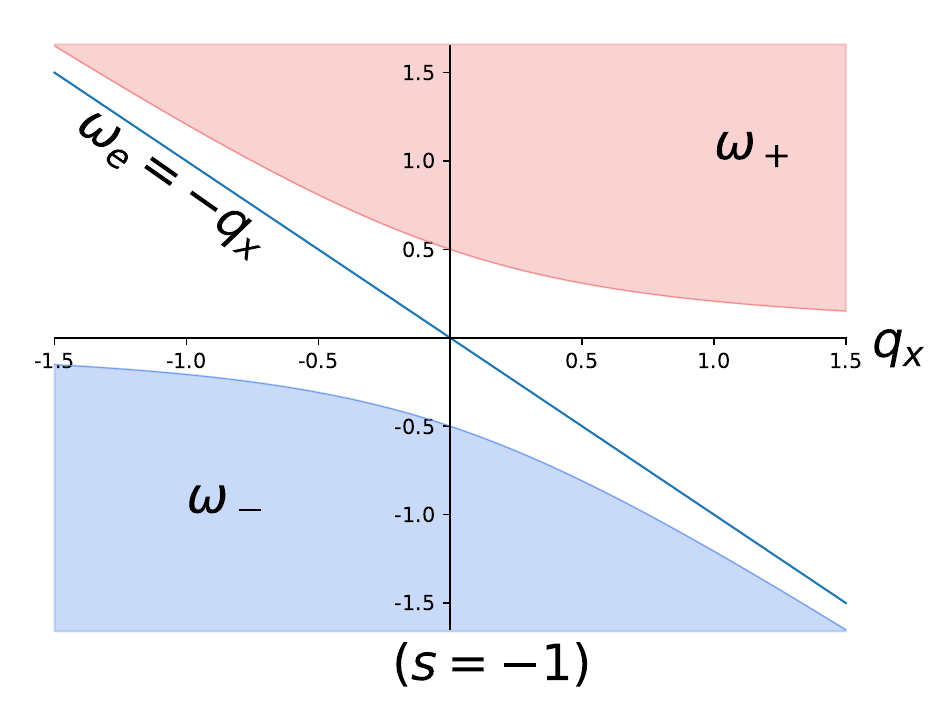}

\caption{Band structure near outer equator (a) for $s =1$ (b) for $s=-1$.}
\label{fig:dispersion_outer}
\end{figure}

The bulk band structure for $s=\pm 1$ are shown in Fig.~\ref{fig:dispersion_outer}(a) and Fig.~\ref{fig:dispersion_outer}(b) respectively. The chiral edge state with dispersion $\omega_e = s q_x$ traverses the bulk gap and remains separated from the bulk modes for all finite $q_x$.
\subsection{Inner equator $(\theta = \pi)$}
Here 
\begin{equation}
    m(y) \sim -\frac{y}{R - r}\,.
\end{equation}
Since $R>r$, $m(y<0) >0$ and $m(y>0) < 0$, i.e., $m$ crosses the inner equator as $+\text{ve } \to -\text{ve}$. Hence
\begin{equation}
    \int^y m (y') \, dy' \sim - \infty \qquad \text{as} \qquad |y| \to \infty \,.
\end{equation}
Consequently, $v$ is square-integrable while $\delta \rho$ diverges as one goes far away from the inner equator on either side. Again, for square integrable bases of the kernels, $\text{ker}_{\mathcal{L}_A} = 0; \,\, \text{ker}_{\mathcal{L}_B} \neq 0$\,.
%
%
Then for pure kernel state, $\omega_e \neq s q_x$, or, $\delta\rho_{\rm edge} = 0$. Then from~\eqref{eq:diffop_A},  $\omega_e=0$ (Goldstone modes). In the laboratory frame these states reduce to purely advected waves
\begin{equation}
    \partial_t v_{\rm edge} + \bar \lambda \partial_x v_{\rm edge}  = 0 \quad  \Rightarrow \quad v_{\rm edge} \sim e^{-i \bar \lambda t } \,.
\end{equation}
Finally, computing the integral
\begin{equation}
    \int_0^y m  \, dy' = r \int_\pi^\theta \frac{\sin \theta \,d\theta}{\varepsilon + \cos \theta}  = - r \ln \left( \frac{\varepsilon -1}{\varepsilon +  \cos \theta}\right)   
\end{equation}
we obtain the topological edge state at the outer equator as
\begin{equation}
    |\delta \Psi_{\rm edge} \rangle = \begin{bmatrix}
         0\\ v_{\rm edge} 
    \end{bmatrix}
\end{equation}
with
\begin{equation}
    v_{edge} = \left( \frac{\varepsilon -1}{\varepsilon +  \cos \theta}\right) \sum_{n\ge 0} [b_n e^{in (\phi - \bar \lambda t)} + \text{c.c.}]\,.
\end{equation}
Note that the envelope of the edge state at the inner equator does not depend on $\eta$ unlike the edge states at the outer equator. So, the inner mode's shape is fixed by the curvature alone, not by any of the material parameters. 

\begin{figure}[t]
\centering

\includegraphics[width=0.48\columnwidth]{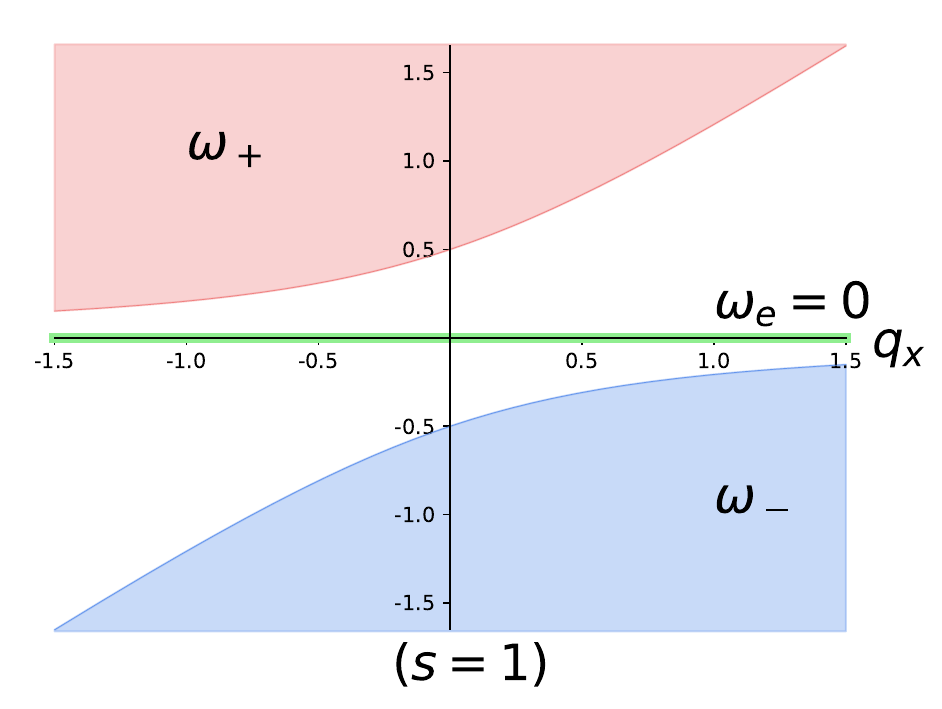}
\hfill
\includegraphics[width=0.48\columnwidth]{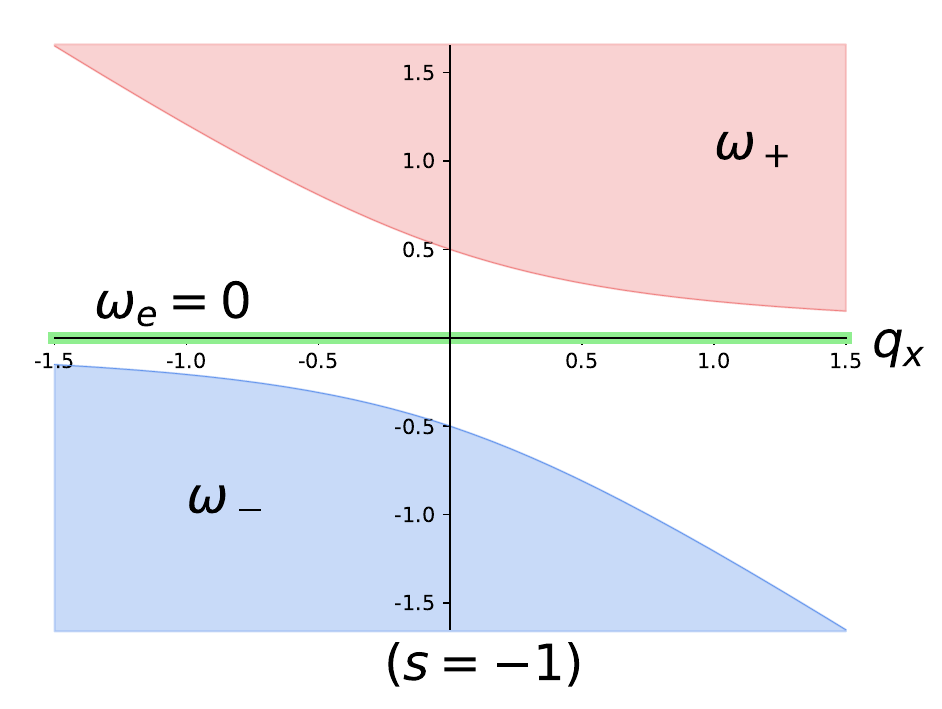}

\caption{Band structure near inner equator (a) for $s =1$ (b) for $s=-1$.}
\label{fig:dispersion_inner}
\end{figure}

In Fig.~\ref{fig:dispersion_inner}(a) and Fig.~\ref{fig:dispersion_inner}(b) it is shown that at the with respect to the inner equator, although the bulk spectra remain same as that for the outer equator, the edge branch now appears to be horizontal line $\omega_e = 0$ (in the comoving frame) lying entirely within the bulk gap. 
\section{Non-Hermitian Index Theorem}
The channel selection described above is not just a choice but it is dictated by topological index theorems. For $q_x = 0$, the Dirac operator~\eqref{eq:Dq} exhibits chiral structure, since it anticommutes with $\sigma_z$. Moreover, in the adiabatic limit $\mu \to \bar v_1 \,  (\mu \neq 0)$, the off-diagonal blocks are adjoints of one another. 
\begin{equation}
    D_0  =  \begin{bmatrix}
         0 &  Q^\dagger \\ Q & 0 
    \end{bmatrix} ; \qquad Q = - i\sqrt{\bar v_1} (\partial_y + m)
\end{equation}
The general case $(\mu \neq \bar v_1)$ is homotopic to it with a continuous homotopy 
\begin{equation}
    Q_t = -i \sqrt{\bar v_1} \partial_y - i \left[ (1-t) (\mu/\sqrt{\bar v_1}) + t \sqrt{\bar v_1} \right] m 
\end{equation}
where $t \in [0,1]$. Note that, during this interpolation $m(+\infty) \neq 0$, and $m(-\infty) \neq 0$, i.e., the bulk spectrum remains gapped. Since the Fredholm index is invariant under homotopy, the general non-Hermitian Dirac operator has exactly the same index as $D_0$ has.

As seen in the previous section, at most one of the two kernels $\text{ker }Q$ and $\text{ker }Q^\dagger$ can contain a normalizable state. Therefore, there exists a one-dimensional Callias index~\cite{callias1978}
\begin{align}
    \text{ind}(Q) &= \text{dim ker }Q - \text{dim ker }Q^\dagger \nonumber \\
                  &= \frac{1}{2} [\text{sgn }m(+\infty) - \text{sgn }m(-\infty) ]
\end{align}
This immediately gives on the outer equator $\text{ind}(Q) = + 1$, and whereas, it is $-1$ on the inner equator. Since, the Callias index is the graded count of localized edge excitations, $\text{ind}(Q) = n_{\rm density} - n_{\rm Goldstone}$, a positive index signifies an excess of density edge modes over Goldstone edge modes, while a negative index signifies an excess of Goldstone edge modes over density edge modes. Thus, the dispersion relations we considered in the previous section, is not merely a choice, but dictated by the index theorem.

The above is a global index theorem. However, combining it with the identity~\eqref{eq:riccati}, we obtain a local index 
\begin{equation}
    \text{ind}(Q(\theta_0)) = \text{sgn } K_G (\theta_0)\,.
\end{equation}
It says the surface with positive curvature binds a density mode, whereas that with negative curvature binds a Goldstone mode.

Now, turning $q_x$ back on promote the isolated zero mode at $q_x =0 $ to deform continuously into a dispersing edge branch, where the eigenfrequency varies continuously with $q_x$. As $q_x$ is varied, the branch crosses the bulk gap exactly once. The spectral flow (SF) of the one-parameter family $\{ D(q_x) \}$ is defined as
\begin{equation}
    \rm SF = N_{\uparrow} - N_{\downarrow} = \text{ind}(Q)
\end{equation}
where $N_{\uparrow}$ and $N_{\downarrow}$ count eigenvalues crossing zero upward and downward respectively. By bulk-edge correspondence, the jump of the bulk Chern number
\begin{equation}
    \Delta C_+ = s\, \rm SF = s \,\text{ind}(Q)\,.
\end{equation}
Now, summing the local index over the surface links it to the Euler characteristics. Finally, summing over all domain the Poincar\'e-Hopf index theorem reads 
\begin{equation}
    \sum_{\rm walls} \text{ind}(Q(\theta_0)) = \sum_{\rm walls} \text{sgn }K_G = \frac{\chi(T^2)}{2}
\end{equation}
which equates $0$ for the torus. 

The active flock on the torus therefore realizes two complementary index theorems simultaneously, an open one-dimensional Callias index theorem locally at each of the outer and inner equator and a global Poincar\'e-Hopf index theorem globally on the entire compact surface. Thus the spectral theory and the topology manifests the same underlying topological structure of the active polar flocks on a defect-free curved surface.
\section{Discussion}
In this paper, we have shown that an ordered active polar flock on a torus supports topologically sound modes even though non-existence of a defect. This establishes that the topologically protected modes are a property of the local geometry (Gaussian curvature and connection), not of defects or boundaries. We have shown the torus hosts two comoving protected channels of opposite character on it, a density edge state on the positively curved outer equator and a Goldstone edge on the negatively curved inner equator, selected by a local index theorem. The global is fixed by the Euler characteristic.

The two channels transports at different speeds. One of them modulates density and the other only orientation. Remarkably, their existence and localization depend only on a single parameter, the aspect ratio of the torus $R/r$. On a fat ring $(R \gg r)$, the two channels are sharply localized, while in the thin-ring limit $(R \ll r)$, the gap closes and the modes delocalize as expected for an active polar flock on a cylinder.
This work indicates that there exists a non-Hermitian Callias-type index theorem in the hydrodynamic limit. 
\acknowledgments
PD acknowledges financial support from the Institute Seed Grant, National Institute of Technology Karnataka (NITK) Surathkal, India.

\end{document}